\documentclass{iopart}
\usepackage{iopams}
\usepackage{graphicx}
\usepackage{units}
\usepackage{subfig}
\usepackage{ulem}
\usepackage{color}
\usepackage[latin1]{inputenc}
\usepackage{cite}

\begin{document}

\title{Optical Absorption Measurements on Crystalline Silicon Test Masses at \unit[1550]{nm}}

\author{Jessica Steinlechner, Christoph Kr\"uger, Nico Lastzka, Sebastian Steinlechner, Alexander Khalaidovski and Roman Schnabel}

\address{Institut f\"ur Gravitationsphysik,
Leibniz Universit\"at Hannover and Max-Planck-Institut f\"ur
Gravitationsphysik (Albert-Einstein-Institut), Callinstr. 38,
30167 Hannover, Germany}

\ead{roman.schnabel@aei.mpg.de}

\date{\today}

\begin{abstract}
Crystalline silicon is currently being discussed as test-mass material for future generations of gravitational wave detectors that will operate at cryogenic temperatures. We present optical absorption measurements on a large-dimension sample of crystalline silicon at a wavelength of \unit[1550]{nm} at room temperature. The absorption was measured in a high intensity monolithic cavity setup using the photo-thermal self-phase modulation technique. The result for the absorption coefficient of this sample with a specific resistivity of $\unit[11]{k\Omega cm}$ was measured to be $\alpha_{\rm A}=\unit[(264 \pm 39)]{ppm/cm}$ for an intensity of 700 W/cm$^2$. 
\end{abstract}
%
%
\section{Introduction}

The initial (1st) and advanced (2nd) generations of interferometric gravitational wave (GW) detectors employ suspended fused silica test masses and use a laser wavelength of \unit[1064]{nm}~\cite{ligo,geo1,virgo}. Observatories beyond the 2nd generation will require very high laser powers to reduce the quantum noise at frequencies above $\unit[\sim50]{Hz}$, while at lower frequencies it is promising to cool the test masses to cryogenic temperatures to reduce thermal noise.

While at room temperature fused silica shows a high mechanical Q-factor~\cite{penn2006} and low optical absorption~\cite{hild}, the mechanical Q-factor decreases by several orders of magnitude at cryogenic temperatures~\cite{schwarz2009, schnabel2010}. This makes fused silica unsuitable as test mass material for cryogenically operated GW observatories.
Crystalline silicon, however, shows a promising mechanical Q-factor at room temperature that even increases towards cryogenic temperatures up to $2\times 10^9$~\cite{Nawrodt2008, Guigan1978}. Due to the very high absorption coefficient of about \unit[10]{/cm} at \unit[1064]{nm}, silicon test-masses require a change to higher laser wavelengths, where the absorption coefficient decreases rapidly~\cite{Keevers1995}.

A wavelength of \unit[1550]{nm} is located within the silicon energy gap and appears promising because of the availability of appropriate optical components and lasers developed for the telecommunication sector. Also, strong squeezed-light sources of $>\unit[12]{dB}$ are available at \unit[1550]{nm}~\cite{Mehmet2011} to increase the sensitivity of GW detectors beyond the quantum limit~\cite{Geo2}. An up to now open question are measurements of the optical absorption coefficient at \unit[1550]{nm} in the temperature range from a few kelvin up to 300 K. Unfortunately, the measurements presented in~\cite{Keevers1995} were done at slightly shorter wavelength up to \unit[1450]{nm}. Furthermore, these measurements are a prediction of the absorption coefficient that is based on photo-current measurements. They might not include various effects that lead to optical absorption but do not generate charge carriers. Also the photo-current measurement might not show all absorption effects that are relevant for thermal noise.

In this paper we present direct optical absorption measurements on two silicon samples in large dimensions at room temperature using the photo-thermal self-phase modulation technique~\cite{SHG}. Silicon sample A forms a monolithic cavity. The measurement technique is perfectly adapted to this setup and therefore gives very precise results. While the photo-thermal effect delivers the absorption coefficient, with this method the round-trip loss is measured independently at the same time. Since losses apart from absorption are small for this monolithic setup, the two results confirm each other. Silicon sample B from another manufacturer did not have a dielectric coating and was placed at Brewster's angle in a Fabry-Perot cavity. Being less stable and containing reflection and scattering losses, the measurement with this setup only served as an order of magnitude estimation to yield an approximate upper and lower limits of the absorption. The slightly different setup backed the fact that the optical absorption found for sample A was not coincidentally untypically high or low and was not mainly caused by the dielectric coatings.
%
%
\section{Absorption Measurements on Silicon Sample A}

Sample A was manufactured by \texttt{Siltronic AG}~\cite{SILCHEM} with the Czochralski technique. The crystal's orientation is (111). The material has a specific resistivity of about \unit[11]{k$\Omega$cm}, which indicates a low doping or contamination with foreign atoms. According to the manufacturer it is a low boron doping, which is a p-donator. This means the impurity concentration is approximately $2\times 10^{12}$ atoms per cm$^3$~\cite{Sze}.
Sample A is the purest material available to us in the required dimensions at ordering time.

The substrate was cut and polished into a cylinder with the rotation axis being parallel to the (111) axis. The cylinder's diameter was $2\times R=\unit[10]{cm}$ ($R$ is the substrate radius), the length was $L=\unit[6.5]{cm}$. The end surfaces were polished to be convex curved with a radius of curvature of \unit[1]{m} to form a cavity with a free spectral range (FSR) of \unit[663]{MHz}.

The substrate's curved end surfaces were coated using ion beam sputtering (IBS). The high-reflection coatings consisted of SiO$_2$ and Ta$_2$O$_5$ and had a design reflectivity of $>$ \unit[99.9]{$\%$} at a wavelength of \unit[1550]{nm}. Hence, the coated substrate formed a monolithic cavity with beam propagation along the (111) axis.
%
%
\subsection{Experimental Setup}
\label{sec:monolith}

\begin{figure}
	\centering
		\includegraphics[width=12cm]{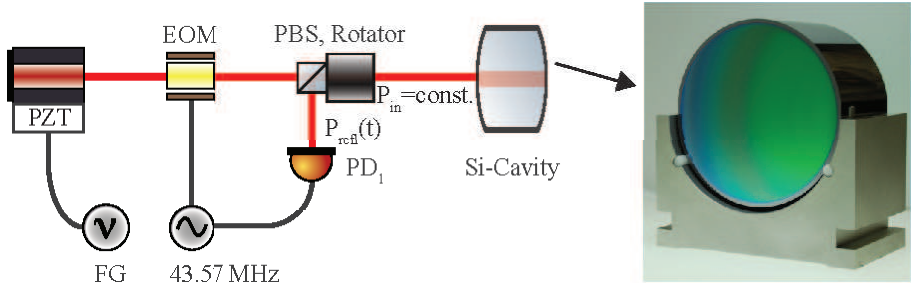}
		\caption{Schematic of the experimental layout. The cylindric silicon substrate with convex curved and coated end surfaces formed a monolithic cavity. A function generator (FG) that actuated the piezo electric transducer (PZT) of the laser modulated the laser wavelength. Photo detector PD$_1$ detected the power reflected by the cavity and showed the cavity resonance peaks. On the right side, a photograph of the monolithic cavity with \unit[6.5]{cm} length and \unit[10]{cm} diameter is shown.}
	\label{fig:experimental_setup}
\end{figure}

Fig.~\ref{fig:experimental_setup} shows a schematic of the experimental setup. A laser beam at a wavelength of \unit[1550]{nm} was mode-matched to the eigenmode of the monolithic cavity. To calibrate the time-axis of our measurements we used frequency markers. An electro optical modulator (EOM) generated these frequency markers by imprinting sidebands at a frequency of \unit[43.57]{MHz} onto the light field.

Photo detector PD$_1$ detected the reflected light, which was separated from the incident field by a combination of a Faraday rotator and a polarizing beam splitter (PBS). Demodulating PD$_1$ and creating a Pound-Drever-Hall-type~\cite{PDH} error-signal generated the frequency markers.

For exploiting the photo-thermal self-phase modulation technique, the laser frequency was scanned around the resonance frequency of the cavity via a piezo electric transducer (PZT). An increasing wavelength corresponded to a shortening of the cavity and a decreasing wavelength to a lengthening of the cavity.

\begin{figure}
	\centering
		\includegraphics[width=11cm]{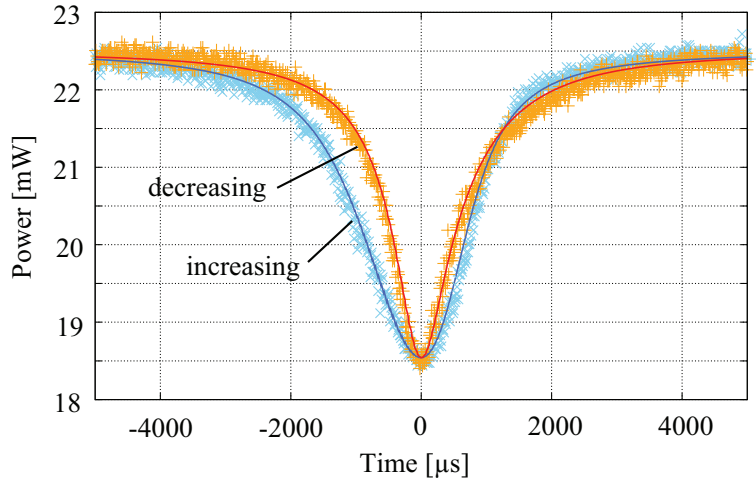}
		\caption{Example of measured (crosses) and simulated (solid lines) reflection peaks at a scan frequency of \unit[0.6]{Hz}:
		The broad (blue) peak forms for an increasing wavelength, the narrow (orange) one for a decreasing wavelength. Without absorption, both peaks would be identical.}
	\label{fig:600mHz}
\end{figure}

The input laser power was \unit[22]{mW} for all measurements resulting in an intensity of $\unit[700]{W/cm^2}$ within the substrate. We performed several measurements by varying the scan frequency starting from \unit[0.2]{Hz} to \unit[49]{Hz} in 18 steps with a constant scan amplitude. The frequency for each measurement number can be found in Tab.~\ref{tab:frequency_measurement_number}. Because the PZT showed a hysteresis, the actual wavelength change had to be calibrated for each frequency for increasing and decreasing wavelength. The time axis of each measurement was calibrated from the scan frequency, the FSR, and the detected error signal. In Figure~\ref{fig:600mHz}, an example measurement for a scan frequency of \unit[0.6]{Hz} is shown. This frequency corresponds to a scan velocity of about \unit[2]{ms/peak} at full width half maximum (FWHM). The yellow crosses (narrow peak) show the measured peak for a decreasing wavelength with the corresponding simulation in red (solid line). The light blue crosses (broad peak) show the peak for an increasing wavelength with the simulation in dark blue (solid line). Without absorption, the two peaks were identical.

\begin{table}
 	\centering
  	\caption{Scan-frequencies used for the single measurements.}
 		\label{tab:frequency_measurement_number}
 				\begin{tabular}{llllllllllllllllllll}
  			\hline
        Measurement number 	&1	&2	&3	&4	&5	&6	&7	&8	&9\\
 				Scan frequency [Hz] &0.2&0.4&0.6&0.8&1.0&1.2&1.4&1.6&1.8\\
 				\hline
 				Measurement number 	&10	&11	&12	&13	&14	&15	&16	&17	&18\\
 				Scan frequency [Hz] &2	&4	&6	&8	&10	&20	&30	&40	&49\\
 				\hline
 				\end{tabular}
\end{table}

%
%
\subsection{Measurement Analysis and Results}

To calculate the absorption coefficient $\alpha$ from the measured peaks as shown in Fig.~\ref{fig:600mHz}, the peaks were fitted. For the fitting process we used the parameters from Table~\ref{tab:monolith_silicon_parameters} as well as the in-coupling reflectivity R$_1$, the effective out-coupling reflectivity $\widetilde{R_2}$, and for measurements with a visible thermal effect $\alpha$ as fitting parameters. Here, $\widetilde{R_2}$ is the effective reflection of the out-coupling coating accounting for the entire cavity round trip loss apart from the transmission of the in-coupling coating. The values for material parameters were taken from literature~\cite{frey06,web,komma}. For the geometric parameters, we used values based on our best knowledge of the cavity design. A Nelder-Mead algorithm was run to find the best set of fitting parameters minimizing the standard deviation of measurement and simulation. 13 of 18 single measurements showed a visible thermal effect and were used to derive the absorption. The remaining five measurements showed no thermal effect due to the high scan-frequency. \unit[8]{Hz} was the threshold above which no thermal effect occurred. All 18 measurements were used to derive R$_1$ and $\widetilde{R_2}$. The results obtained from measurements without thermal effect were consistent with the remaining results.

\begin{figure}
	\centering
		\includegraphics[width=12cm]{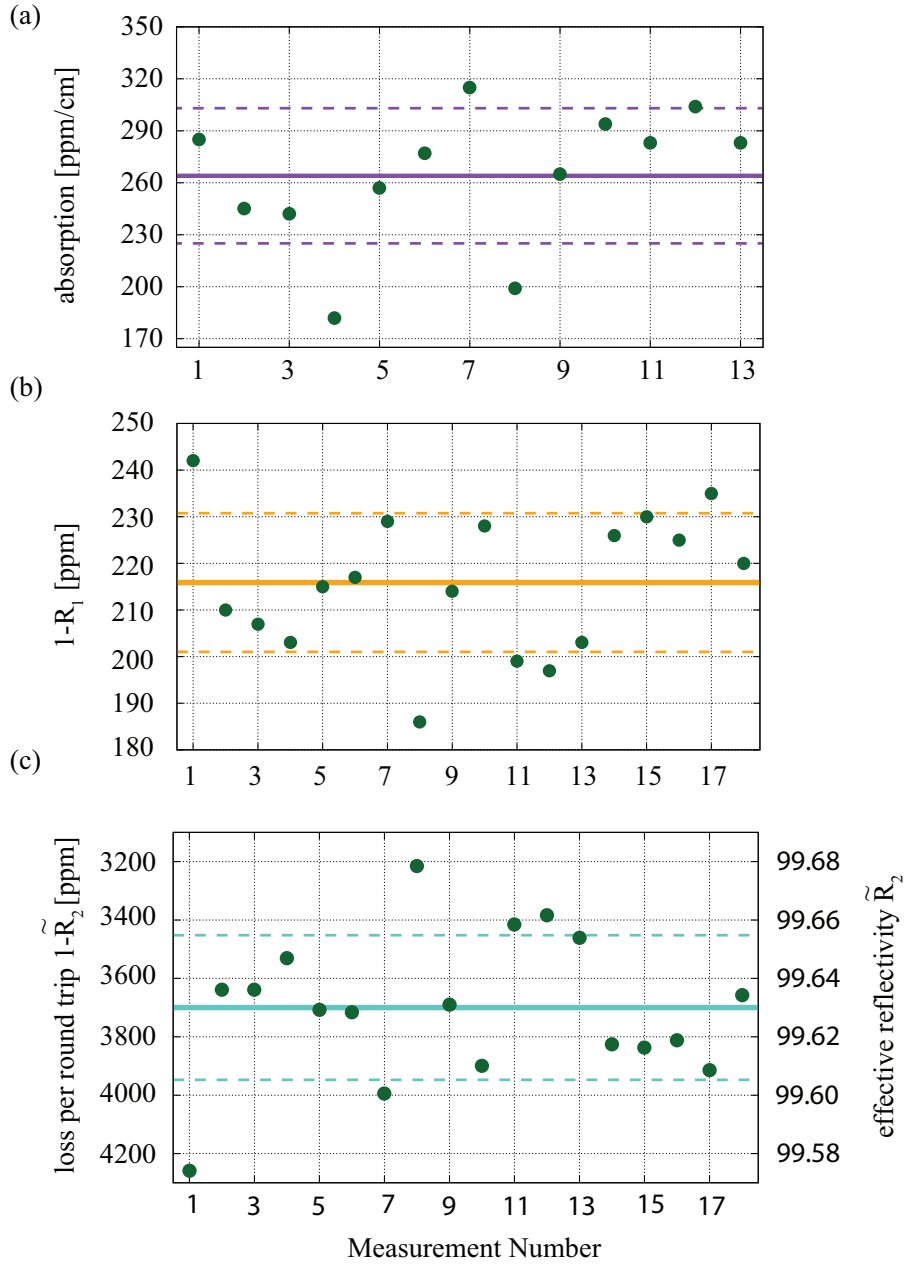}
		\caption{(a) Results for the absorption from single measurements (green dots). The mean value and the standard deviation of $\alpha=\unit[(264 \pm 39)]{ppm/cm}$ are given by the purple line and the dashed purple lines, respectively. The dots in (b) and (c) show the results for the in-coupling reflectivity R$_1$ and the effective out-coupling reflectivity $\widetilde{R_2}$ with $R_1=\unit[(99.9784 \pm 0.0015)]{\%}$ in orange and $\widetilde{R_2}=\unit[(99.630 \pm 0.025)]{\%}$ in turquoise.}
	\label{fig:results}
\end{figure}

Figure~\ref{fig:results}(a) shows the results for the absorption coefficient $\alpha$ derived from the 13 different measurements (dark-green dots). The purple lines show the mean value of all single results and their standard deviation which is $\alpha=\unit[(264\pm 39)]{ppm/cm}$ ($\unit[39]{ppm} \; \widehat{=} \; \unit[15]{\%}$).

\begin{table}
 	\centering
  	\caption{Material and geometric parameters of the monolithic silicon cavity used for the simulations.}
 		\label{tab:monolith_silicon_parameters}
 				\begin{tabular}{lll}
       	\hline
        Geometric parameters 							& 																	&Source\\
  			\hline
        crystal radius $R$ 								& $\unit[5]{cm}$										&specified by manufacturer\\
 				crystal length $L$ 								& $\unit[6.5]{cm}$									&specified by manufacturer\\
 				beam waist $\omega_0$ 						& \unit[217.8]{$\mu$m}							&calculated from $L$, $n$ and $ROC$\\
 				\hline
        Material parameters 							& 																	&\\
 				\hline
 				index of refraction $n$						& $3.48$ 														&\cite{frey06} \\
				thermal expansion $a_{\rm th}$ 		& $\unit[2.62\times10^{-6}/]{K}$ 	&\cite{web}\\
        thermal refr. coeff. d$n/$d$T$ 		& $\unit[1.87\times10^{-4}/]{K}$ 		&\cite{komma}\\
        specific heat $c$ 								& $\unit[714]{J/(kg\,K)}$ 					&\cite{web}\\
        density $\rho$ 										& $\unit[2330]{kg/m^{3}}$ 					&\cite{web}\\
        thermal conductivity $k_{\rm th}$ & $\unit[140] {W/(m\,K)}$ 					&\cite{web}\\
 				\hline
 				\end{tabular}
\end{table}

The results for the (power) reflection R$_1$ of the in-coupling coating are shown in Figure~\ref{fig:results}(b). The mean value and standard deviation of all 18 single measurements is $R_1=\unit[(99.9784 \pm 0.0015)]{\%}$. Measurement numbers 14--18 did, as stated above, not exhibit a visible thermal effect.

Figure~\ref{fig:results}(c) shows the results for $\widetilde{R_2}$ for all single measurements (dark-green dots) and their mean value and standard deviation (turquoise lines) of $\widetilde{R_2}=\unit[(99.630 \pm 0.025)]{\%}$. This result was used to cross-check the obtained absorption values. An absorption of $\alpha_{\rm{total}}=\unit[0.343]{\%}$ per round-trip results in an effective absorption coefficient of $\alpha=\unit[(264 \pm 39)]{ppm/cm}$ taking into consideration the round-trip length of \unit[13]{cm}. The round-trip loss added to $\widetilde{R_2}$ results in a new effective reflection $\widetilde{R_2}'=\widetilde{R_2}+\alpha_{\rm{total}}=\unit[99.973]{\%}$ that still contains the cavity scattering loss. Since the pure reflection can be assumed to be very similar to $R_1$ (identical coating design, but different coating runs) the results for the three parameters agree perfectly. This is an additional consistency check and not an automatical consequence of the simulation.

We repeated the series of measurements for a polarization rotated by 90$^\circ$. Further we exchanged in-coupling and out-coupling coating and repeated the measurements for the two polarizations. As expected, in each case the results for the absorption coefficient agreed with the result presented above within the error bars.
%
%
\subsection{Error Propagation}

The error bar of \unit[39]{ppm} or $\unit[15]{\%}$ corresponds to the standard deviation of 13 independent measurements using different scan-velocities. An additional error bar arises from uncertainties in the simulation input parameters. To estimate this error, we individually changed the input parameters listed in Table~\ref{tab:monolith_silicon_parameters} by $\pm\unit[10]{\%}$ and recalculated R$_1$, $\widetilde{R_2}$ and $\alpha$.
We found that for most parameters the influence on $\alpha$ is approximately linear and none of the changed parameters caused a change of the result for $\alpha$ by more than $\unit[15]{\%}$.

For $n$, the error bar of the value from literature is in the order of $10^{-5}$~\cite{frey06} and therefore negligible, the uncertainty of ${\rm d}n/{\rm d}T$ is in the same order of magnitude~\cite{komma}. $a_{\rm th}$ and ${\rm d}n/{\rm d}T$ affect the result as a sum. Since $a_{\rm th} \ll {\rm d}n/{\rm d}T$, an uncertainty of $a_{\rm th}$ is negligible. The error bars of $c$, $\rho$ and $k_{\rm th}$ are not known to us. We estimate that the uncertainties of the cavity geometric input parameters as well as for the calibration of the time axis and in the measurement of the mode-matching are below \unit[10]{\%}.

If the simulation input parameters are precise within $\unit[10]{\%}$, our statistical error bar of $\unit[15]{\%}$ is the dominating error contribution.
%
%
\section{Measurements on Silicon Sample B}

The absorption measured with sample A was unexpectedly high (see discussion). To verify that this sample did not absorb untypically much, or that the coating did not cause the absorption, a second experiment with a different sample B was performed. The two samples differed in manufacturer and crystal orientation.

A cylindric substrate with one inch diameter and \unit[3.5]{cm} length was manufactured by \texttt{Mateck}~\cite{Mateck} in (100) orientation using the Czochralski procedure. The (100) axis is the rotation axis of the cylinder. The cylinder end surfaces were polished to be parallel with a tolerance of $< \unit[30]{''}$.
The material was declared by the manufacturer to be undoped with a resistivity of $> \unit[5]{k\Omega cm}$.

\begin{figure}
	\centering
		\includegraphics[width=8cm]{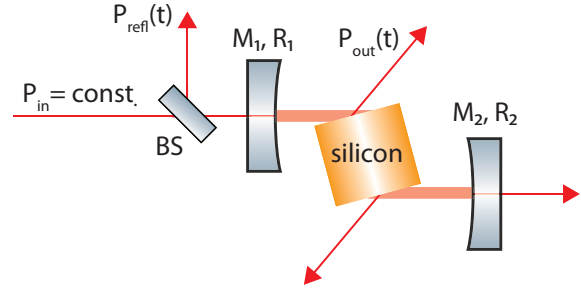}
		\caption{Experimental setup for the measurement on the second silicon sample. The beam was coupled into the cavity through mirror M$_1$. The silicon substrate was placed in the cavity at Brewster's angle of \unit[74]{$^\circ$} which caused a strong beam displacement. M$_2$ was the end mirror of the cavity. The resonance peaks were detected in reflection of the beam splitter (BS). The laser wavelength was scanned with a function generator.}
	\label{fig:silicon_extern}
\end{figure}

The substrate was placed in a resonator at Brewster's angle of \unit[74]{$^\circ$} to minimize round trip loss due to reflection at the substrate surface (see Figure~\ref{fig:silicon_extern}). The mirrors were clamped to an aluminium spacer. The laser beam coupled into the cavity through the in-coupling mirror M$_1$. The reflected beam power P$_{\rm refl}$ was detected in reflection of the beam splitter (BS). (Using the BS instead of the Faraday rotator and PBS combination in the first setup, does not change anything for the experiment.) The modulation of the laser, the calibration of the time axis and the measurement procedure were identical to the procedures described in Section~\ref{sec:monolith} for the monolithic cavity setup.

Entering the substrate at Brewster's angle causes an elliptical beam profile within the substrate. Since not necessary for other experiments, a non-circular beam profile is not implemented in our simulation program. The discussed measurement had the goal of independently determining a lower limit for the optical absorption to confront the obtained value with the results from sample A. Therefore, a calculation of an upper limit for the power density and thus for the heat distribution within the substrate is sufficient. This is given by a circular beam profile with the radius of the minor semi-axis of the elliptical profile. Since a higher power density requires a smaller absorption to cause the same thermal effect, this assumption yields a lower limit for the absorption coefficient.

\begin{figure}
	\centering
		\includegraphics{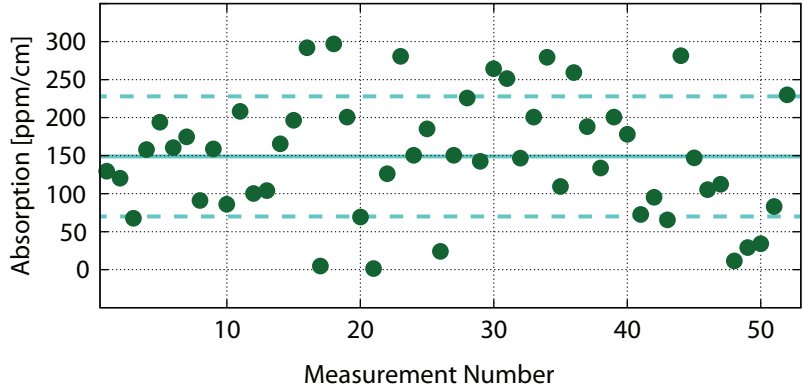}
		\caption{The results for the lower limit of the absorption coefficient of sample B obtained using photo-thermal self-phase modulation are shown by the green dots. The light blue line illustrates the mean value of all single measurements of $\alpha_{\rm B}=\unit[149]{ppm/cm}$ and the dashed light-blue line the error bars of \unit[79]{ppm/cm}.}
	\label{fig:extern_thermal}
\end{figure}

From 52 single measurements, the lower limit for the absorption was found to be $\alpha_{\rm B}=\unit[(149 \pm 79)]{ppm/cm}$. The results of the single measurements are shown in Figure~\ref{fig:extern_thermal} (green dots). The mean value of all measurements with the standard deviation are depicted by the light-blue line and the light-blue dashed lines, respectively. Despite a shared spacer for mirrors and substrate, the external cavity setup proved to be instable and prone to acoustical disturbances, which caused large error bars due to the statistical fluctuations of the detected peaks.

The error propagation was already discussed for sample A. The uncertainty of the result is dominated by the large standard deviation, while the errors in the material parameters are negligible in first order approximation.

Apart from the elliptical beam profile, the large statistical error shows that the second setup is much more instable and therefore disadvantageous compared to the monolithic setup. Nevertheless, the lower limit for the absorption of $\alpha_{\rm B}=\unit[(149 \pm 79)]{ppm/cm}$ allows the conclusion that the absorption in sample A did not primarily originate in the dielectric coatings.

In a second series of measurements, the round trip loss of the cavity was measured. This provided an the upper limit for the absorption coefficient. For this measurement, two mirrors with identical coatings were used and the round-trip loss was minimized by inclining and rotating the substrate. The maximum impedance matching was found to be $\unit[(21.2 \pm 0.3)]{\%}$. (An impedance matching of $\unit[100]{\%}$ means that the reflected power at resonance is zero.)
Using the design reflectivity of R$_1$=R$_2=\unit[(99.97 \pm 0.01)]{\%}$ the optical loss was calculated to L$_{\rm RT}=\unit[(4400 \pm 1200)]{ppm}$. L$_{\rm RT}$ contains the entire optical loss that consists of absorption of and scattering at the mirror coatings as well as of the reflection and scattering $P_{\rm out}$ at the silicon substrate surface. The latter occurs twice per round trip because of entering and leaving the substrate, respectively. These reflections are caused by non-perfect plan-parallel end surfaces of the substrate, the wave-front distortion of the beam, scattering and limitations in fine-adjustment. The laser beam passes $2\times \unit[3.5]{cm}/{\rm cos}(90^\circ-74^\circ)=\unit[7.28]{cm}$ of the substrate per round trip. This results in a loss of $\unit[(4400 \pm 1200)]{ppm}/\unit[7.28]{cm}=\unit[(604 \pm 165)]{ppm/cm}$ and forms the upper limit for the absorption coefficient.

Nevertheless, the result of $\unit[70]{ppm/cm} \leq \alpha \leq \unit[770]{ppm/cm}$ for silicon sample B suggests that the result obtained with sample A was typical for samples of the degree of purity involved.
%
%
\section{Discussion}

\begin{figure}
	\centering
		\includegraphics[width=10cm]{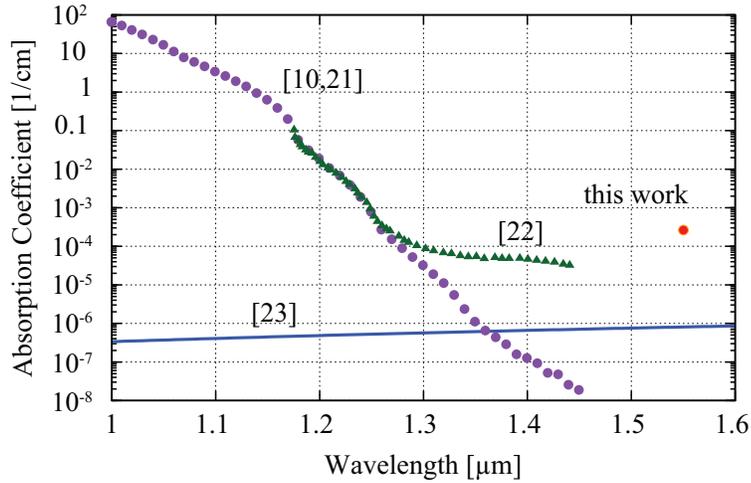}
		\caption{The purple dots and green triangles show band-band absorption values $\alpha_{\rm BB}$ of silicon from literature~\cite{Keevers1995, Green} (purple dots) and~\cite{ana} (green triangles). The free carrier absorption $\alpha_{\rm FC}$ for a p-doping of $N=\unit[2\times 10^{12}]{/cm^3}$ (sample A) was calculated using~\cite{Soref1986} (blue line). The red dot shows our result for sample A.}
	\label{fig:absorption_literature}
\end{figure}

Fig.~\ref{fig:absorption_literature} shows the absorption result from the present work in comparison to earlier absorption results from spectral response measurements on solar cells taken from literature (purple dots and green triangles)~\cite{Green,ana} and to a theoretical prediction of the absorption caused by the residual boron contamination of sample A~\cite{Soref1986}.

Previous measurements~\cite{Green,ana} found a much lower absorption of silicon than measured here. However, these results are also not consistent. Although no measurements at \unit[1550]{nm} are available, the absorption measurements from Keevers and Green predict an absorption coefficient of $\alpha<\unit[0.02]{ppm/cm}$ at \unit[1550]{nm}~\cite{Keevers1995}, while the measurements from Anagnostopoulos predict an absorption coefficient in the order of $\alpha\approx \unit[50]{ppm/cm}$~\cite{ana}. Green and Keevers explain their much lower absorption results by suggesting contamination of Anagnostopoulos' sample, due to doping or unintended foreign atoms. The assumed kind of contamination is, however, not specified in either of the two publications. The band-band absorption $\alpha_{\rm BB}$ is the lower absorption limit for intrinsic crystalline silicon. Preliminary results obtained by Degallaix et al.~\cite{degallaix} point to a light-intensity dependent absorption in silicon. An estimation based on the power density in our setup showed that our result is not significantly influenced by that effect.

Even very low doping or contamination of the silicon samples can dominate the band-band absorption in the infrared region due to the free carrier absorption $\alpha_{\rm FC}$~\cite{Soref1986}. Since the semiconductor industry is by far the largest area of application for crystalline silicon, samples almost always are doped or at least slightly contaminated with the doping material that is normally used in the apparatus for the crystal growth. A specific resistivity of $\unit[11]{\rm k\Omega cm}$, which is the specific resistivity of sample A, corresponds to a p-doping of $N=\unit[2 \times 10^{12}]{/cm^3}$ (N is the number of doping atoms). The number of doping atoms was calculated from the specific resistivity using~\cite{Sze}. The blue line in fig.~\ref{fig:absorption_literature} shows $\alpha_{\rm FC}$ for a p-doping of $N=\unit[2 \times 10^{12}]{/cm^3}$ (blue line) calculated using~\cite{Soref1986}. $\alpha_{\rm FC}$ and the number of doping atoms are proportional~\cite{Schmid}. This theory predicts an absorption coefficient of $\alpha\approx \unit[1]{ppm/cm}$ for sample A, which is two orders of magnitude below our measurement results. Therefore, this doping theory does not explain the absorption of samples A and B. However, the theories for absorption due to doping generally are optimized for much higher doping than in our case. It is therefore possible that the prediction for residual-doping-induced absorption is not accurate, and further theoretical and experimental investigations are required to clarify this issue. Another possible and plausible explanation for the deviation from the numbers published by Keevers and Green is that their photo-current measurements did not include effects that, while leading to optical absorption, would not lead to generation of carrier charges. Such photo-current measurements are thus not unconditionally transferable to optical absorption and may rather be used to derive a lower limit. Finally, the specific resistivity values provided by the manufacturers not necessarily provide information about all kinds of contamination present in the crystal and thus can not be used as an absolute measure for the crystal's purity. Further investigation of the residual contamination is required to derive a model for the dependence of the optical absorption on residual doping and contaminations, which are not reflected in the specific resistivity value.
%
%
\section{Conclusion}

We measured the room temperature absorption coefficient of two silicon samples A and B which differed in manufacturer and crystal orientation. According to the manufacturer, sample A had a residual contamination by boron atoms with $N=\unit[2\times 10^{12}]{/cm^3}$. For sample B, the specific resistivity was specified to be $< \unit[5]{k\Omega cm}$, this corresponds to a p-doping of $N<\unit[4\times 10^{12}]{/cm^3}$.

For sample A, a result of $\alpha_{\rm A}=\unit[(264 \pm 39)]{ppm/cm}$ was obtained. The consistent results for optical absorption and round trip loss induce that no other process apart from the optical absorption provides a significant loss contribution.

For sample B, upper and lower limits of $\unit[70]{ppm/cm} \leq \alpha_{\rm B} \leq \unit[770]{ppm/cm}$ were derived for the absorption coefficient. This confirmed the result for sample A not to be extraordinarily high or low. Since uncoated, the measurement of sample B additionally proved that the absorption of sample A did not originate in the dielectric coatings but was due to absorption in the bulk substrate and/or the surface oxide layers. An identification of the origin of the optical test mass absorption will be subject to further measurements that shall be conducted in the near future.
%
%
\section*{Acknowledgements}

We acknowledge support from the SFB/Transregio 7, the International Max Planck Research School (IMPRS) on Gravitational Wave Astronomy, and from QUEST, the centre for Quantum Engineering and Space-Time Research.
%
%

\end{document}